# Hardware System Implementation for Human Detection using HOG and SVM Algorithm

Van-Cam Nguyen, Hong-Tuan-Dinh Le, Huu-Thuan Huynh

Computer and Embedded System Laboratory

University of science, VNU-HCM, Ho Chi Minh City, Vietnam

*Abstract*— Human detection is a popular issue and has been widely used in many applications. However, including complexities in computation, leading to the human detection system implemented hardly in real-time applications. This paper presents the architecture of hardware, a human detection system that was simulated in the ModelSim tool. As a co-processor, this system was built to off-load to Central Processor Unit (CPU) and speed up the computation timing. The 130x66 RGB pixels of static input image attracted features and classify by using the Histogram of Oriented Gradient (HOG) algorithm and Support Vector Machine (SVM) algorithm, respectively. As a result, the accuracy rate of this system reaches 84.35 percent. And the timing for detection decreases to 0.757 ms at 50MHz frequency (54 times faster when this system was implemented in software by using the Matlab tool).

## I. INTRODUCTION

Nowadays, human detection is an attractive title for researchers in the computer vision field around the world. Those systems can be handled by specific co-processors, which off-load for the CPU and improve the speed of the detection process. In this paper, the architecture of a human detection system in hardware was proposed. The organization of this paper is as follows. In Section II, some of the researched results, are related to object detection. In Section III, the architecture of the system is described in an overview. In Section III, the architecture of the system is described in an overview. The components of the system are presented in detail in Section IV. Measurement results of the system are demonstrated in Section V. Last, conclusions are given in Section VI.

## II. RELATED WORK

Many researchers have proposed diverse techniques to deal with object detection [1]–[7]. Generally, object detection includes two significant processes, are feature extraction and classification. For the present, the HOG algorithm is proven to offer better feature extraction that could significantly outperform existing feature sets for object detection [1], particularly with regard to pedestrians. [2] proposed several methods to simplify the computation, such as the conversion of the division and square root, and implemented the proposed architecture on FPGA to achieve real-time requirements. [3] proposed human detection. However, these are computationally complicated and not in detail for full processes: extracting, training, and detecting. The proposed architecture in this paper is a full human detection system in both software (for training) and hardware (for extracting and detecting). Some of the methods were used in this system based on [4]-[7].

## III. OVERVIEW THE SYSTEM

The human detection system includes two parts: the software was used in the Matlab tool to train the input dataset. And the hardware was simulated in the ModelSim tool, which detected a 130x66 pixels RGB input image.

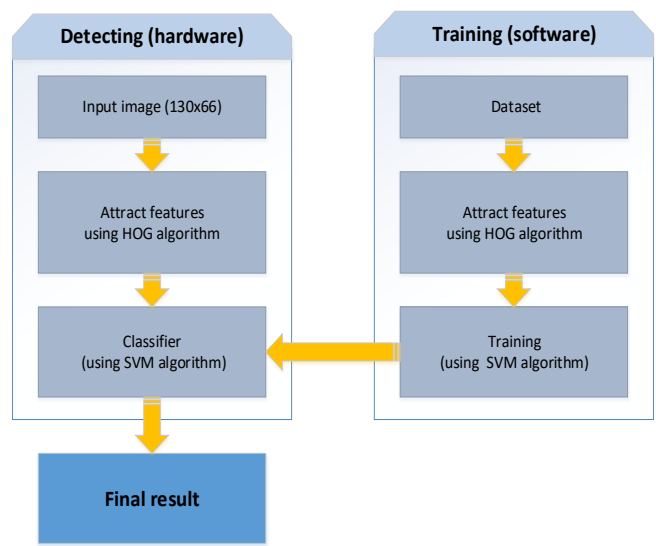

Fig.1. The diagram of the full human detection system, implemented in the Matlab tool (for attracting, training and detecting).

*Training:* From the input dataset for training, the HOG characteristics of each image (gray image, 8-bits) were attracted. These features would be trained by using the SVM algorithm to classify humans. This process was implemented on the Matlab tool.

*Detecting:* The input image would be grayscale (8 bits), extracted by the HOG algorithm. These characteristics were combined with the previously trained parameters (from training) to classify a person (equal to 1) or no person (equal to zero) according to the SVM algorithm. This process was implemented on hardware by using the ModelSim tool for simulating.

## IV. SYSTEM IN DETAIL

### A. Training

The hyperplane would be used to classify the input image. The training block used two algorithms, the Histogram of Oriented Gradient (HOG) algorithm, to extract characteristics of the RGB input image. And the SVM (Support Vector Machine) algorithm, is applied to get the characteristic parameters for the hyperplane. Corresponding two sub-units are shown in Fig. 2 and Fig. 5 below.

*1) The first sub-unit : the extractor.*

The Histogram of Oriented Gradient (HOG) algorithm shown in Fig. 2, comprises of 6 stages

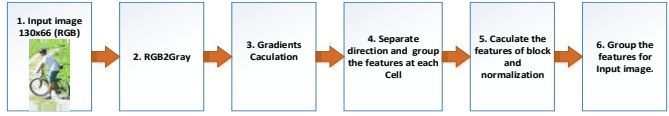

Fig. 2.    The diagram of the extractor of an image on MATLAB tool.

In the 1st stage, the dataset for training includes 2,795 negative images and 4,202 positive images, taken from INRIA and MIT standard datasets. This dataset contained quite a few instances of person and not for training. The 2nd stage for color standardization, which converted the input color image (RGB) from the dataset to grayscale (8-bits) to extract. This process aimed to reduce the space of memory for containing. The algorithm converted the pixel intensity information in Fig. 3(a) to gradient information, as illustrated by the image of gradients for the pedestrian's heel (see Fig. 3.(b)). Gradient parameters are present both in magnitude and direction.

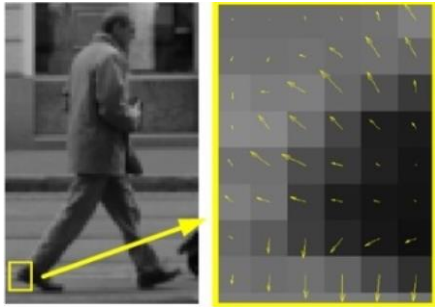

Fig. 3.    (a) Image of Intensities. (b) Image of Gradients.

In the HOG feature extractor, the input image was divided into small parts called cells, normally 8x8 pixels as shown in Fig. 2. Then the gradient in both horizontal and vertical directions was calculated for each pixel within the cell.

$$f_x(x, y) = f(x + 1, y) - f(x - 1, y) \quad (1)$$

$$f_y(x, y) = f(x, y + 1) - f(x, y - 1) \quad (2)$$

Once gradients for both x and y directions were calculated, the gradient magnitude m(x, y) and gradient direction $\theta$ (x, y) could be computed as:

$$m(x,y) = \sqrt{f_x^2(x,y) + f_y^2(x,y)} \quad (3)$$

$$\theta(x,y) = \arctan(\frac{f_x(x,y)}{f_y(x,y)}) \quad (4)$$

For this system, the detection window is fixed at 130x66 pixels and is divided up into small 8x8 pixel spatial regions or cells, which consist of histograms of gradient directions (edge orientations) over the 64 pixels of the cell as shown in Fig. 4. The gradient of each pixel relative to its surrounding pixels was calculated in the 3rd stage, which involved the calculation of the magnitude of each x-axis and y-axis gradient pair and the addition of the result to the relevant cell bin. Over an 8x8 pixel cell, a single 9-element vector was produced which was referred to as the histogram of gradients. In the 4th stage, blocks were generated by locally normalizing groups of four cells i.e. 2x2 cells, in order to improve the invariance to illumination and shadowing. The resultant block vector had 36 elements. Collation of the blocks over the full detection window (7x15 blocks) was carried out in the 5th stage to produce HOG descriptors. In the final stage, a Support Vector Machine (SVM) classifier received the 3780 (7x15x36) vectors and multiplied them with its set weights to achieve the human detection chain.

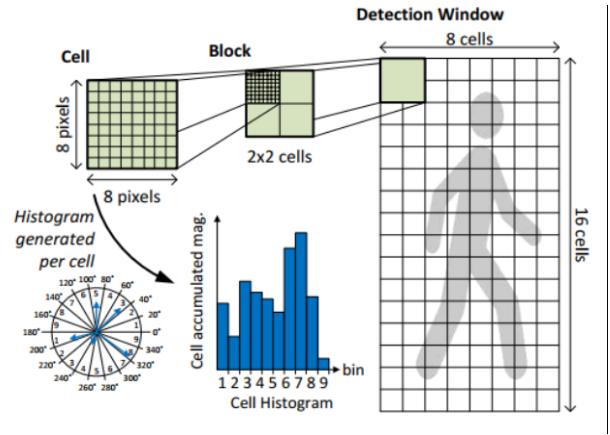

Fig.4.    The detection window divided into blocks and cells producing an array of histograms per window.

At the final stage, a histogram normalization calculation is generated by combining all histograms belonging to one block, which consists of four cells. The normalized result can be realized as:

$$v_i^n = \frac{v_i}{\sqrt{\|v\|_2^2 + \varepsilon^2}} \quad (5)$$

where $i$ is a number from 1 to 36 (four cells x nine bins), $v_i$ is the vector corresponding to a combined histogram for a block region, $\varepsilon$ is a small constant to avoid dividing to zero.

*2) The second sub-unit : the classifier.*

The output of the dataset for training on each cell is a strong classifier and the strong classifiers from all the cells of an input image patch are combined to construct a feature vector. A linear SVM classifier [7] is then trained over the feature vectors for the final classification. The procedure of SVM training for classification is summarized as follows.

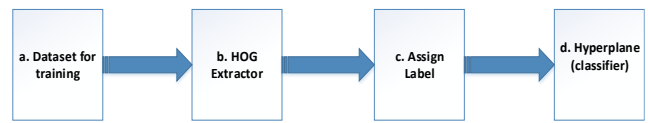

Fig. 5.    The diagram of searching hyperplane for classification on MATLAB tool.

The 130x66 pixels RGB images from the dataset (a) were taken in turn to extract the characteristic at (b) (using the HOG algorithm). Based on the characteristics of the input images, proceed to assign the label to each corresponding image (label 1, corresponding to the image of person, label 0 corresponds to the image without person) at (c). Then, at (d), these were used to get the hyperplane for classification, based on the SVM algorithm. The characteristic expression for that hyperplane is:

$$D(x) = \mathbf{W}.\mathbf{X} + b \qquad (6)$$

Vector $\mathbf{W}$ and bias $b$ are two characteristic parameters for the hyperplane, respectively, which has been trained on Matlab tool using the SVM algorithm. Vector $\mathbf{X}$ is characteristic for input image, which will be detected. The hyperplane delimiter plays an important role in classifying it, deciding which class a dataset belongs to. To perform the classification, the SVM simply determines whether a set of data lies to which side of the metaphor separated by the expression.

$$D(X) = sign\,(\mathbf{W}.\mathbf{X} + b) \qquad (7)$$

Equation (7) is a function for the sign of $\mathbf{W}.\mathbf{X} + b$. If D $(\mathbf{X}) < 0$ then the data set will be located below the hyperplane delimiter. Else is D $(\mathbf{X}) = 0$ then the data set will be in the superimposed binomial. Else, D $> 0$, the data set will be located above the superimposed plane.

*B. Detecting*

The parameters, which were trained in the software, then proceed to design the hardware for the 130x66 input image. This system would be used the ModelSim tool to test the ability and development to build the system's hardware on FPGA board:

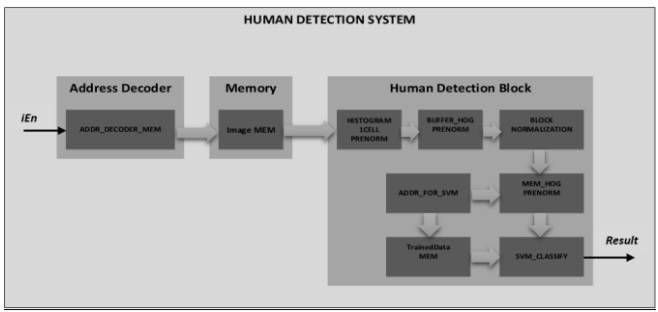

Fig. 6. The diagram of human detection system in hardware, implemented in ModelSim tool (for detecting).

The system began to perform image extraction and image identification if the *iEn* signal was positive (set to 1). When the *iE*n signal was positive, the *ADDR_DECODER_MEM* block would begin address decoding of each corresponding 8-bits grayscale cell contained in the *Image MEM*. After every 108 clock cycles, the *ADDR_DECODER_MEM* block would be resolved once (each cell is extracted in 108 cycles, the next cell values are added once the cell was extracted). From the input addresses, the *Image MEM* block would outlast the pixel values corresponding to those addresses, providing data for the *HISTOGRAM_1CELL_PRENORM* block that extracted the HOG attribute for the input cell. The features after being calculated would be saved in block *BUFFER_HOG_PRENORM*. These features were then included in *BLOCK_NORMALIZATION* to standardize. This block completed the normalization after 47 clock cycles. The final characteristic value was the normalized value. *BUFFER_HOG* was a block containing the standardized character of the input image to be detected. The result would be the output of the *SVMCLASSIFY* block. This block was based on the data provided by the *BUFFER_HOG* block (containing the input image characteristics) and the *TrainedData_MEM* block (containing data trained on Matlab tool), using the SVM algorithm.

The floating-point IEEE 754 32-bits was used to minimize errors in the calculation process. And, in this system, the CORDIC algorithm was applied to calculate magnitude and direction (angle) based on two initial values X and Y.

The COordinate Rotation DIgital Computer (CORDIC) is often used to calculate trigonometric values. Calculations in trigonometry are often complex. A hardware system to calculate trigonometric functions, it will take a huge resource to build and the computation process is very complex. Therefore, the CORDIC algorithm is a good choice to approximate the trigonometric values when they are designed in the hardware.

In this system, the CORDIC algorithm was used to approximate *arctan* function and magnitude of a value. Fig. 7 shows in detail how the CORDIC algorithm works in this design.

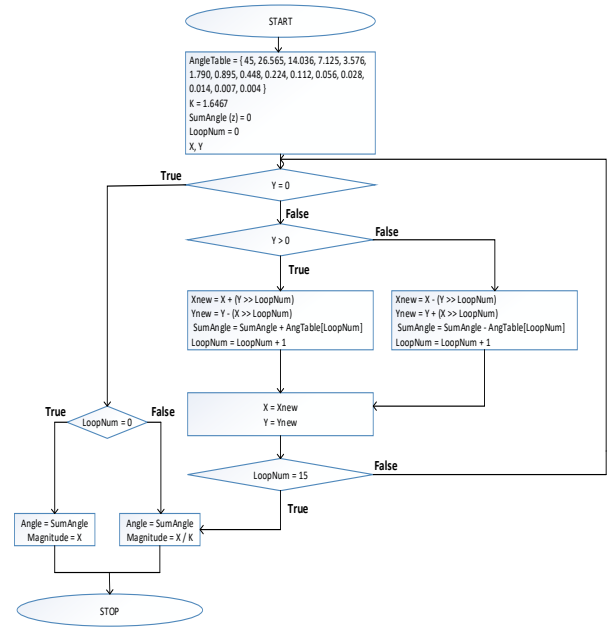

Fig. 7. The flowchart of CORDIC algorithm in hardware.

The hardware design blocks the gradient (including magnitude and angle) of a pixel, used the CORDIC algorithm.

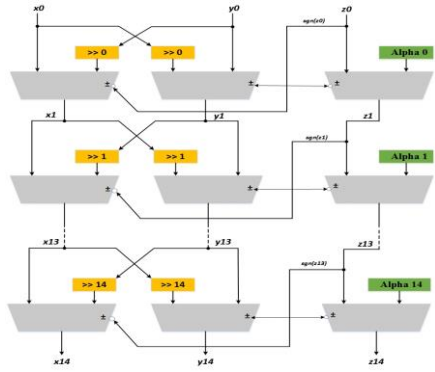

Fig. 8. The CORDIC block diagram in hardware.

Calculating up to n = 14 (ie. up to 15 angle values from the Lookup Table are retrieved). The *Block_NormalizationCore* was used the *Newton-Raphson* algorithm to approximate the square root of the sum of the squares of input bin values. It was shown in detail as [3].

## V. MEASURED RESULT

The human detection system was built in both software (for training and testing) and hardware (for detecting). The measured result is shown as:

### A. *The result of human detection system, implemented in Matlab tool, is shown below:*

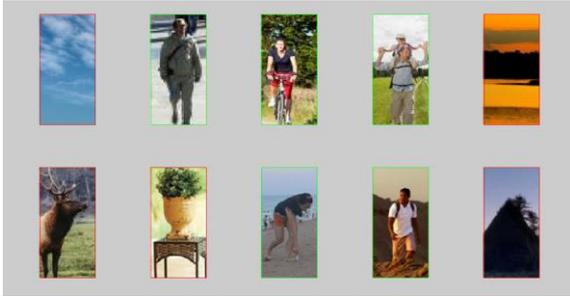

Fig. 9. The green and red rectangles present for images with and without person, respectively.

### B. *The result of human detection system, implemented in wave by using ModelSim tool, is shown below:*

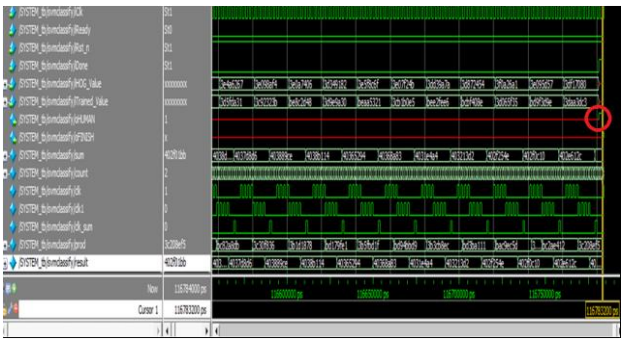

Fig. 10. The result in ModelSim tool of the proposed system.

### C. *The accuracy rate*

The accuracy rate of the proposed system is described in below table. Totally, there were 294 images (includes 160 images with person and 134 images without person, respectively) were tested. And the result reached to 84.35% for accuracy rate.

TABLE I
The accuracy of the proposed human detection system

| Input images | True detection | False detection | Accuracy rate |
|---|---|---|---|
| With person | 134/160 | 26/160 | 83.75% |
| Without person | 114/134 | 20/134 | 85.07% |
| Total | 248/294 | 46/294 | 84.35% |

### D. *The timing*

The timing of the human detection system is shown in detail below. The time for detection decreased to 0.757 ms at 50MHz frequency in ModelSim tool (54 times faster when this system was implemented in software by using Matlab tool).

TABLE II
The timing of the proposed human detection system

| Timing | In Matlab tool | In ModelSim tool |
|---|---|---|
| For training | 298.323 s | None |
| For attracting | 16 ms | 0.411 ms |
| For detecting | 41 ms | 0.757 ms |

## VI. CONCLUSIONS

This system had proposed a complete method of detecting human in static images. The approach of research was based on ideas in image processing, computer vision and machine learning, using the Histogram of Oriented Gradients (HOG) features combined with Support Vector Machines (SVM) algorithm on FPGA board in static images. This paper illustrated the architecture of the human detection system, trained the dataset was available on Matlab tool, which will be an important basis for the detection process later. And, this paper designed the core of the system on both hardware and software. However, it was not possible to detect human in different resolutions, only the 130x66 pixels images. The system has not yet completed to load onto the real board. The accuracy rate was not high (84.35%). The hardware detection rate used the SVM algorithm after training is comparable to the performance of the software identification. The accuracy rate depends on the dataset for training when the training dataset is not sufficient to be detected human.

*For future development*

Building hardware blocks will be combined with Qsys tool to complete the human detection block (with a grayscale input image, size 130x66) on hardware. Then, will be combined with the ARM-core chip to extract the input image from the camera to provide the detection block. And the ARM-core chip will be used to follow the localization of the human detection block. Eventually display the recognition result on the screen.

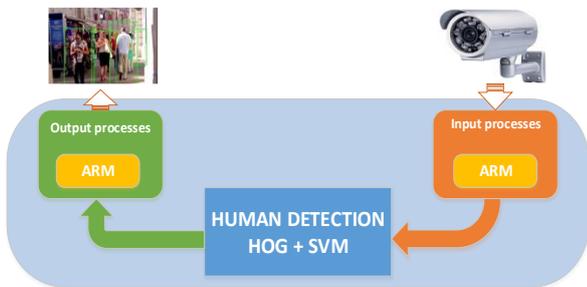

Fig. 11. The completed human detection system in the future.

ACKNOWLEDGMENT

This work has been undertaken in Computer and Embedded System Laboratory (CESLAB), University of Science, Vietnam National University, Ho Chi Minh City, Vietnam.

APPENDIX

Available code at:
https://github.com/VanCamNguyen/Human-Detection